# Self-Supervised Convolutional Audio Models are Flexible Acoustic Feature Learners: A Domain Specificity and Transfer-Learning Study

Mattson Ogg

Research and Exploratory Development Department
Johns Hopkins University Applied Physics Laboratory
Laurel, USA
mattson.ogg@jhuapl.edu

*Abstract*— Self-supervised learning (SSL) leverages large-scale unlabeled audio data to pre-train representations that generalize well to new tasks. Historically, SSL models have been developed independently for speech and non-speech applications. This report examines the domain specificity of a self-supervised convolutional (BYOL-A) model's pre-training data relative to tasks in different downstream domains (e.g., speech or non-speech tasks). Regardless of whether they were pre-trained on speech, non-speech, or both types of audio, BYOL-A models learned features that supported strong performance on most downstream tasks, including those outside their explicit pre-training domain. Only minor performance differences were observed based on the domain specificity of the pre-training datasets. Representational similarity analysis was used to directly compare these models' representational spaces both to one another and to a set of well-studied acoustic dimensions. Despite differences in initial pre-training data, models trained with this SSL approach converged on similar representational spaces, characterized by sensitivity to spectral content in different frequency bands, spectral variability, and pitch. Together, these results demonstrate that SSL methods can learn flexible representations without labels, supporting transfer learning, fine-tuning, or data exploration applications when the downstream data are similar, but also (at least for this variety of SSL convolutional model) when there may be a domain mismatch. These results invite further study around the importance of inductive biases in different SSL pre-training strategies and the potential generalizability of different kinds of pre-training data. In sum, this work offers new insights into SSL pre-training across matched and mismatched audio domains, insights into the resulting model representations and cross-domain generalization.

## I. INTRODUCTION

For most real world audio applications (including human language technologies, medical diagnostics, and bioacoustics) audio data is trivial to collect but time consuming to annotate. This creates a bottleneck for traditional supervised machine learning methods that rely on accurate labels to map input data to the appropriate outputs. A common workaround has been to leverage a model that has been previously trained for a different, often more general task, and then to fine-tune or adapt that model to a new, more specific task [1]. The goal is to leverage the knowledge a model has learned on one task and repurpose it for (or transfer it to) a different task or dataset. Usually this form of transfer-learning requires less labeled data for the new task (i.e., the target domain) to re-map the model's representation to the new application than supervised models trained from scratch [2]. Notable examples of transfer learning for audio tasks include using models trained for speaker recognition to classify whether a person has Alzheimer's [3] or Parkinson's [4] disease from their speech, or using models trained to recognize sound events from audio data in YouTube videos [5] (or other data [6]) to support the recognition of bioacoustic information. However, there are open empirical questions with respect to how to optimize this transfer process given the data used to initially pre-train a model and the downstream tasks it might be applied to later. Domain specificity refers to the extent to which a model benefits from pre-training on data similar to its downstream tasks. A mismatch between the data used to pre-train a model and the data used for fine-tuning or evaluation (often called a domain shift) can greatly reduce the performance of audio models [7].

SSL has become a prominent approach for pre-training models, enabling more efficient transfer learning. These models aim to learn representations without any labels at all (e.g., [8]). These methods generate robust, generalizable representations by identifying invariant patterns or salient similarities (and differences) among pre-training examples. Because these techniques don't require labels, they can easily ingest huge quantities of data, allowing for a massive increase in the scale of pre-training.

Self-supervised methods have achieved impressive results for speech and non-speech tasks, but these lines of research have often proceeded separately. For example, pre-training transformers to predict masked portions of their input over huge quantities of speech data supports the successful transfer of the model to new, unseen, languages [9] or to new audio conditions [9,10]. Experiments that study the transfer of models pre-trained on speech data to downstream non-speech tasks have been limited. Similarly, some approaches for non-speech tasks try to create representations that are robust against modifications applied to training examples such as added noise, reverb or time-frequency manipulations [11,12]. These perform well on many non-speech benchmarks and have even shown some capability for performing simple speech tasks downstream, but the space of different input pre-training datasets (sometimes called training diets) has not been well explored.

In summary, self-supervised pre-training methods and transfer learning are effective and may provide a scalable and adaptable platform for future research and development but open questions remain regarding how SSL models deal with potential domain-shifts during transfer learning. This work aimed to understand how pre-training on speech or non-speech audio diets of comparable volumes affects downstream transfer to speech tasks, non-speech tasks, or speech and non-speech tasks (i.e., voice activity detection or VAD). Follow-up analyses quantified how different training diets influenced the features each SSL model learned.

## II. METHODS

### A. Pre-Training Data and Processing

Approximately 5000 hours of speech [13-20] and non-speech data [21-28] was collected from standard public datasets to pre-train the SSL models (see Table I). Volumes of the speech and non-speech data were closely matched, with the non-speech data having a slightly longer total duration and the speech data containing slightly more examples. These involved a variety of common speech (near- and far-field, broadcast, English and non-English) and non-speech (discrete audio events, soundscapes) conditions.

In general, all audio was originally recorded from natural sources. Where necessary, the data were re-sampled to 16kHz. Only the pre-specified training partitions of each dataset were used where indicated. Some datasets contained long, continuous audio recordings. In these cases, each audio file was subsampled into 10-second clips (with no overlap).

TABLE I. PRE-TRAINING CORPORA

| Corpus | Audio Domain | Duration (Hours) | Examples |
|---|---|---|---|
| freefield1010 | Non-Speech | 21.4 | 7,690 |
| WHAM 48k Noise^ | Non-Speech | 56.6 | 20,363 |
| SONYC-UST | Non-Speech | 37.6 | 13,538 |
| MUSAN^ | Non-Speech | 106.5 | 38,343 |
| FSD Superset* | Non-Speech | 190.2 | 72,449 |
| Audioset | Non-Speech | 4,817.2 | 1,749,124 |
| TED-LIUM^$ | Speech | 1,193.5 | 429,644 |
| LibriSpeech | Speech | 897.0 | 263,095 |
| CHiME-5^ | Speech | 1,111.9 | 400,280 |
| VoxCeleb | Speech | 308.3 | 134,000 |
| Mozilla Commonvoice% | Speech | 692.4 | 435,946 |
| AMI^ | Speech | 792.1 | 285,143 |

^ Sub-sampled to non-overlapping 10-second segments
\* FSD50k plus non-overlapping examples from noisy18k + Kaggle2019 & TEDx Multi-Lingual plus v2
% Obtained from https://github.com/SeanNaren/deepspeech.pytorch (En only)

### B. Self-Supervised Learning

The BYOL-A[1] self-supervised learning framework [11,12] was used to train convolutional neural network models for these experiments. The reader is referred to the original reports for full details, but briefly, this approach adapts the "bootstrap your own latent" approach for learning image representations via data augmentations [29] to operate over audio by converting the acoustic time series to a spectrogram. This method applies different transformations to each input example and trains a 2-dimensional convolutional neural network (CNN) to produce a representation of the input audio that is invariant to these manipulations. BYOL-A applies transformations such as adding noise, cropping, resizing, time-frequency stretching, and amplitude modulation. Convolutional models like these have been shown to be effective for speech (e.g., [30]) and non-speech (e.g., [1,31]) tasks.

Three different versions of the SSL BYOL-A CNN model architecture were trained from scratch using the different pre-training datasets. The CNN model comprises two convolutional blocks (each 3x3 with 64 channels, batch-norm, ReLU activation and 2x2 max-pooling with a stride of 2), two MLP layers (each with 2048 units, ReLU activation and a dropout layer between them), a concatenation layer and temporal (mean+max) pooling. One model was trained on the speech corpora, one model was trained on the non-speech corpora, and the last model was trained on all the data (both speech and non-speech corpora). Each model was trained for 100 epochs. Training was carried out using an Nvidia A100 GPU and took approximately 1.5 (for the speech and non-speech only models) to 3 days (for the model trained on all data). Mean and standard deviation time-frequency statistics of 10,000 random samples from the training data of each model were used for front-end spectrogram normalization. Mean and standard deviation statistics were derived from each model's training data and were not recalibrated for the transfer learning testing datasets, approximating inference conditions for unseen data. Checkpoints for the three BYOL-A models trained in these experiments are available in an online repository[2].

### C. Transfer Learning Experiments

A collection of smaller datasets was held-out for testing the performance of the different models. These consisted of standard benchmark datasets that, together, would characterize a model's performance on different downstream audio domains: speech (limited vocabulary speech recognition, speaker-recognition), non-speech (sound event recognition or SER, instrument recognition), and a task that combines speech and non-speech targets (voice activity detection, or VAD) to gauge performance for domains that may or may not be similar to what a given model was pre-trained on. Models pre-trained on speech data were expected to do well on speech tasks, and models pre-trained on non-speech data were expected to do better on non-speech tasks. The model trained on both datasets was expected to do fairly well across tasks albeit with slightly worse performance than the models trained solely on speech and non-speech data for those matched domains (due to its capacity being the same as those models, but having to represent a wider array of data during training), but this model was expected do slightly better than single-domain models on the VAD task.

---
[1] https://github.com/nttcslab/byol-a
[2] https://github.com/mogg64/byola_domainXfer

All transfer learning experiments were carried out using the same parameters except where noted below. All held-out data were resampled to 16kHz. These experiments adhered to any pre-specified train, validation and testing partitions where available. If these partitions were not explicitly provided then reproducible portions of these datasets were designated for validation and testing (with the rest used for training), as described below.

For each transfer learning dataset an embedding (averaged over time windows) was extracted from each model for each audio file (thus, one embedding per model per file). During fine-tuning, a simple linear layer was learned between those saved embeddings and a set of output units corresponding to the classes in each task (see below). Training for each experiment ran for 50 epochs using the Adam optimizer with a learning rate of 0.0001 to minimize a cross-entropy loss. Weights were applied to the classes when calculating the loss to help account for any class imbalances. The learning rate was reduced if there was no improvement on the validation loss for 5 epochs. The model with the best performance on the validation partition was retained to evaluate the test partition, and accuracy is reported on the test partition.

Additional details for each of these transfer learning experiments is as follows:

- *ESC-50* [32] (*Non-Speech*, SER): A small SER dataset comprising 50 target classes. Data are organized into five partitions. Partition four was used for validation and partition five was used for testing.
- *UrbanSound8k* [33] (*Non-Speech*, SER): A SER dataset comprising 10 target classes. Data are organized into ten partitions. Partition nine was used for validation and partition 10 was used for testing.
- *Nsynth* [34] (*Non-Speech*, Music Instrument Recognition): A dataset of isolated musical instrument notes played using a variety of methods and dynamics. Models were trained to classify which of the 11 target musical instrument families corresponds to each note using the pre-defined train, validation and test partitions over 25 epochs.
- *Speech Commands v2* [35] (*Speech*, Speech Recognition): A limited vocabulary (35 target words) speech recognition task (i.e., keyword classification) using the dataset's pre-defined partitions.
- *VCTK* [36] (*Speech*, Speaker Recognition): A speech dataset containing 110 different target speakers saying multiple utterances. For each speaker, utterances labelled 25 and below were used for validation, and utterances labelled 26 to 50 were used for testing with the rest used for training.

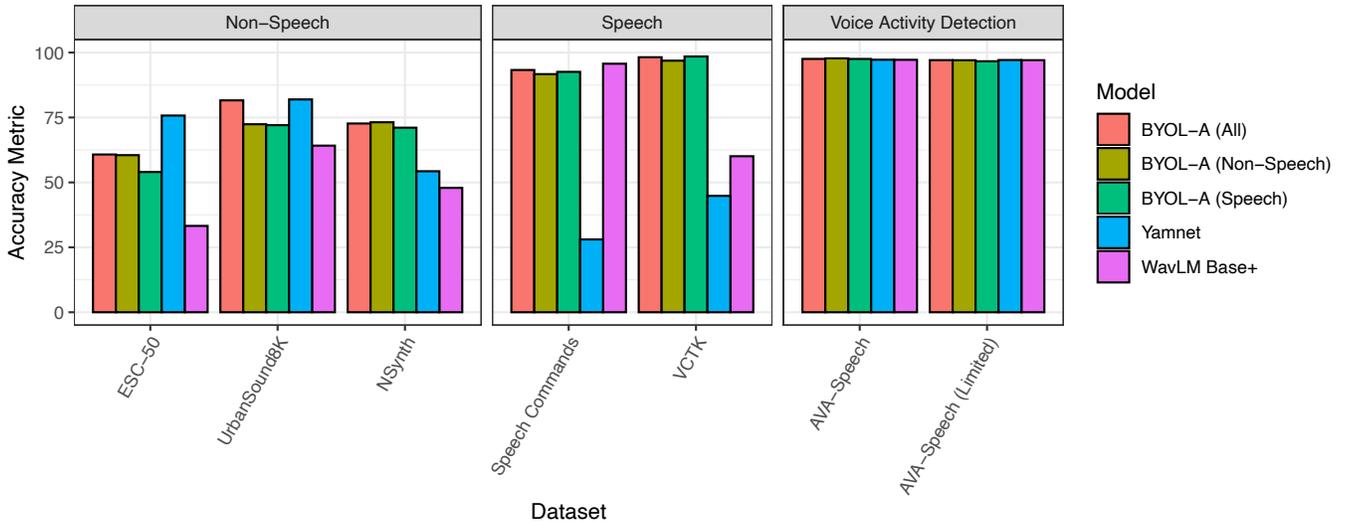

Fig. 1. Model Performance on Transfer-Learning Experiments. *Each dataset was held-out from model training. Performance is reported for each dataset's 'test' partition (see Methods for details). Performance for each model on different downstream tasks is broken out by the different domains of the downstream tasks (speech, non-speech or voice-activity detection). The AVA-Speech voice activity detection experiments use average precision as the performance metric for those analyses.*

- *AVA-Speech* [37] (*Speech and Non-Speech*, VAD): A dataset made up of audio harvested from movies available on YouTube. Audio from 160 of the movies in this dataset was still available (others were removed or are no longer publicly available on YouTube). 15-minute segments of each movie were annotated to indicate time periods where speech was or was not present. Each annotation was parsed into a series of 1-second clips with a 10-millisecond hop. Data were grouped into two classes: "background" (all "NO_SPEECH" annotations) and "speech" (all other annotations, all containing "SPEECH" potentially in the presence of background noise or music). To create partitions, the clips from each movie were sorted by the movie's YouTube ID. Data from the last 10 movies in this list were used for testing and the second to last 10 movies were used for validation. Classes were balanced during training by retaining only 10,000 clips from each class from each film in the training data. Up to 90,000 clips from each film in the testing and validation dataset were used, classes in these evaluation partitions were not balanced. Because of the class imbalance Average Precision was used for assessing performance. When fine-tuning models on this dataset a single output unit with a sigmoid activation was used and fine-tuning comprised 10 epochs.

- *AVA-Speech Limited:* Same as above, but restricted to using only data from the first 30 movies for training.

Finally, to provide context for the performance of the BYOL-A SSL models, these transfer learning experiments were also carried out using popular, high performing domain-specific models that are publicly available. Specifically, this same set of transfer learning experiments was run using embeddings extracted from Yamnet [3] (a model trained to recognize a large number of different non-speech sounds in YouTube videos; [25,31]) and WavLM[4] (the "Base+" model trained using a multi-objective, self-supervised learning approach and a large quantity of English speech data [10]).

*D. Representational Similarity Analysis (RSA)*

A representational similarity analysis [38] was carried out to better understand how the pre-training diet of each model influenced transfer learning performance. This method can help illuminate the underpinnings of model decisions by relating distances among model representations for pairs of stimuli (organized into dissimilarity matrices for each model or model-DSMs) to distances among features that correspond to those stimuli (each organized into acoustic-DSMs per feature). The feature distances that are then most highly correlated with model embedding distances are presumed to be features the model has learned from the pre-training data (Spearman correlation between a given model-DSM and acoustic-DSM). Numerous audio descriptors have been developed [39-42], and previous work has used these to interpret the performance of neural networks on different audio tasks (e.g., [43]).

---
[3] https://www.tensorflow.org/hub/tutorials/yamnet
[4] https://github.com/microsoft/unilm/tree/master/wavlm

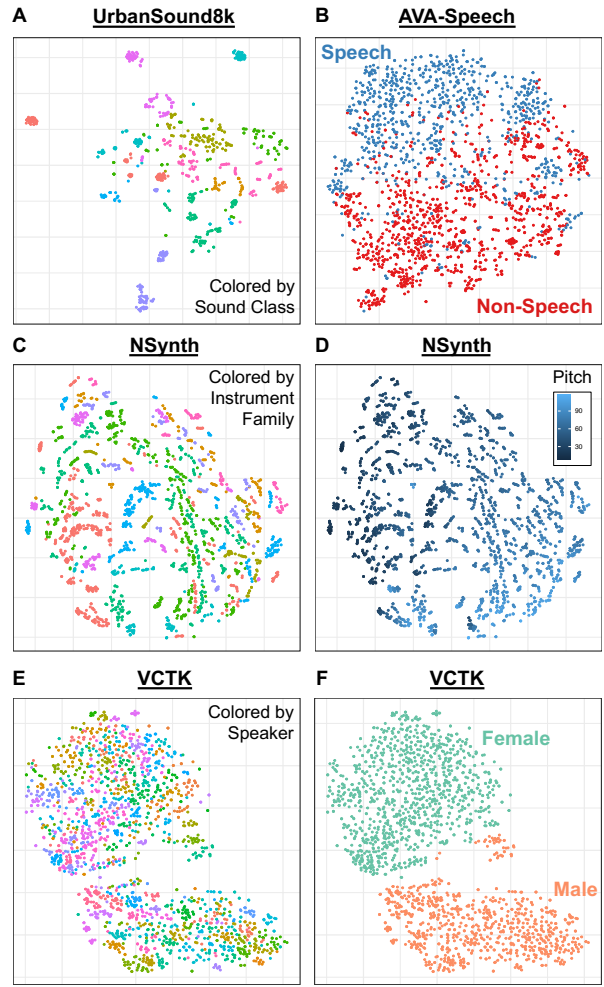

Fig. 2. T-SNE visualization of the BYOL-A (All) embeddings. *Embeddings for up to 2000 random samples from the validation partition of select datasets representing speech, non-speech and VAD audio domains. Axes comprise arbitrary units. Examples are colored by the target classes used in the transfer learning experiments in A, B C and E. The separation of classes in BYOL-A embeddings demonstrates the model's capacity for learning structured and transferable representations. The model also appears to encode pitch (seen in the NSynth data; D) and gender-related information (seen in the VCTK data; F), which is shown by re-coloring the NSynth and VCTK plots relative to these other dimensions. Note axes and orientations are arbitrary for individual T-SNE solutions, which were fit for each dataset individually and are not directly comparable.*

For a subset of files from each held-out dataset's validation partition (up to 1000 files), an embedding was extracted from each model along with a set of acoustic features that have previously been shown to relate to human listeners judgements of speech and non-speech sounds [44]. Cosine distances were calculated among the embeddings for all pairs of these sounds within each dataset along with absolute differences along each feature dimension (excluding identical or redundant pairs). Spearman rank-order correlations were used to measure the similarity among model-DSMs as well as between each model-DSM and the acoustic feature DSMs. Correlations that were interpretable ($r_s > 0$) and survived a false-discovery-rate correction were retained. RSA analyses were carried out for each dataset individually and results were averaged across datasets. Because pitch estimates for many noisy or non-periodic sounds might be unreliable, pitch features were included only for analyses of the NSynth dataset, where each audio example was played at a single pre-determined pitch.

III. RESULTS

Transfer learning experiment results are reported in Figure 1 (see also Supplemental Table 1). The SSL-trained BYOL-A models demonstrated strong performance across tasks, irrespective of their pre-training data or downstream task domains. That is, the pre-training domain only had a small influence on downstream task performance. Only small domain specificity advantages could be observed for the speech and non-speech BYOL-A models. Notably, the BYOL-A model trained on speech and non-speech data demonstrated strong cross-domain robustness, consistently achieving near-peak performance across tasks with diverse input domains.

As expected, the domain-specific models Yamnet and WavLM excelled at their respective speech and non-speech tasks, but tended to perform poorly outside of those domains. Compared to domain-specific models like WavLM and Yamnet, SSL-trained BYOL-A models showed smaller performance reductions for out-of-domain tasks, highlighting their flexibility. All models performed well on the VAD tasks, even in the data-limited regime with only annotations from 30 movies (up to 15 minutes each), which is notable considering this analysis targeted the detection of both clean and noisy speech (i.e., included in the target speech class was speech in the presence of noise or music). However, only the Yamnet model performed well on the very small ESC-50 dataset. Meanwhile, the BYOL-A models performed the best on the music instrument recognition task, even outperforming the Yamnet model which presumably contained pre-training examples with music in them as well as target classes in its supervised pre-training that captured different musical instruments (Gemmeke et al., 2017).

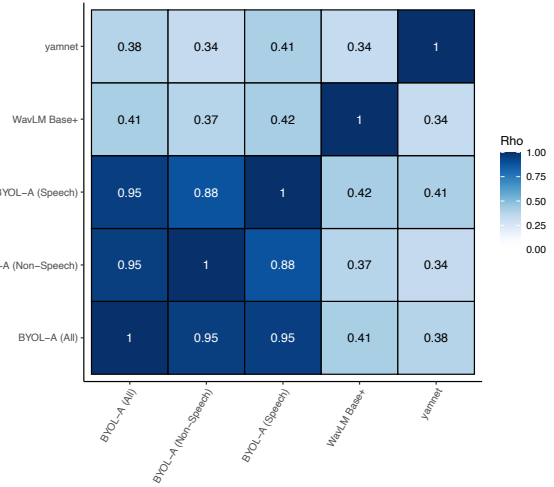

Fig. 4. Inter-model similarity averaged across subsets of the validation corpora used in the transfer learning experiments. *The representations of the SSL BYOL-A models trained in the course of this study were all highly correlated with one another, despite the different domains of audio data used for pre-training. Similar among BYOL-A models was strong than was similarity among models trained with similar domain-specific data. Each cell represents the Spearman correlation between two different model-DSMs (all $p < 0.001$), calculated for each held-out dataset, then averaged across datasets.*

Given the good performance of the BYOL-A models, it was useful to visualize the SSL model's latent space to see how the classes in each dataset were organized (or separated), and if any additional structure was present in the embeddings that might map onto the characteristics of a given dataset. To do this, the embeddings from the model trained on both the speech and non-speech data were projected into two dimensions using T-SNE (see Figure 2). Examples from the validation partitions of representative transfer learning datasets from each domain were visualized in this way (up to 2000 samples, randomly chosen). Since the BYOL-A model's embeddings are not updated after pre-training (just their mapping to the various task-specific output units), this provides a fairly unbiased window into the features learned by these models and how they represent audio events. Class-wise separation is evident across audio domains, with additional structure aligning with acoustic features such as pitch in NSynth (Figure 2D) and gender-related features in VCTK (Figure 2F).

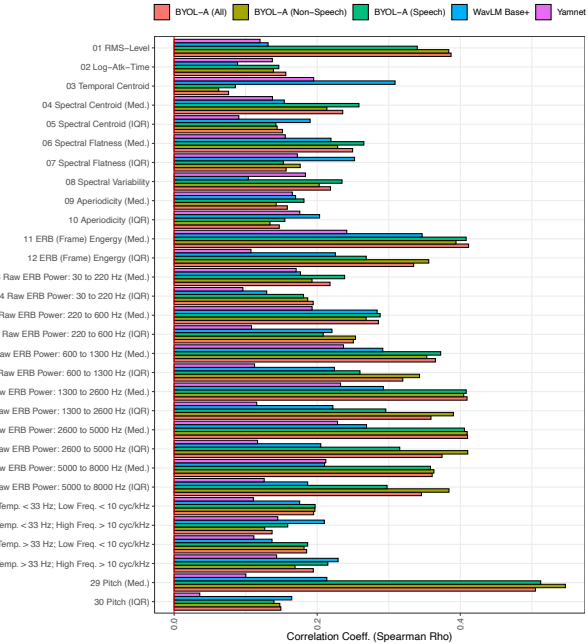

Fig. 3. Representational Similarity Analysis to understand the influence of acoustic features on model representations. *Correlations between pairwise model distances (organized into model-DSMs) and differences along different acoustic feature dimensions (organized into acoustic-DSMs). Associations between Model-DSMs and acoustic-DSMs were obtained via Spearman correlation (all FDR-corrected $p < 0.05$), calculated for each held-out dataset, then averaged across datasets.*

Representational similarity analysis was used to better understand what features the BYOL-A models learned from each pre-training dataset. This method correlated pairwise distances of model embeddings (organized in a model-DSM) for examples from the validation datasets with corresponding pairwise differences along a set of different feature dimensions (organized in feature-wise acoustic-DSMs). This analysis is reported in Figure 3 (see Supplemental Figure 1). Across datasets, BYOL-A models were most sensitive to a sounds' overall amplitude (RMS-level and ERB Energy) especially for frequencies over 600Hz (though the speech-trained model was more particular about which frequencies) and were thus also correlated with the spectral centroid acoustic descriptor. Overall spectral variability was also aligned with BYOL-A representations (as well as within specific frequency bands). Finally, differences in the pitch of NSynth sounds were also correlated with BYOL-A model distances.

The representations learned by all three BYOL-A models were highly similar to one another despite being trained on different domains of sounds (Figure 4; see Supplemental Figure 2). This is apparent through correlations among model-DSMs (correlation among the BYOL-A model distances $r_s$ = 0.88 to 0.95, all $p < 0.001$). This is also apparent in a comparison of how strongly different acoustic features correlated with their embedding distances for a given dataset, the overall distribution of acoustic features that aligned with the BYOL-A models were significantly correlated across models ($r_s$ = 0.93 to 0.98, all $p < 0.001$).

## IV. DISCUSSION

SSL has the potential to revolutionize acoustic machine learning research and enable advances in speech and audio applications. However, open questions remain with respect to how transferable these representations are if they are trained primarily on a specific domain of data and are subsequently used to evaluate or explore another set of data, possibly from a different domain. This work studied one effective method for training self-supervised representations using an augmentation procedure and a 2-dimensional convolutional network (BYOL-A). Results indicated that these models were fairly effective at a wide variety of downstream tasks regardless of the kind of data they were initially pre-trained on. The model trained on both speech and non-speech data was especially effective in some cases and only small domain specificity advantages can be seen for some tasks. Overall, each of the BYOL-A models studied here performed fairly well across downstream tasks. By contrast, the domain specific models Yamnet and WavLM performed very well in their target domains but tended to underperform outside of those.

There are some notable limitations to this work. In particular it will be useful to expand these analyses to involve self-supervised transformer pre-training methods since these are another standard architecture for self-supervised learning and these methods perform very well on downstream speech tasks [10] as well as some non-speech tasks (see [45] for initial work in this area). This leads to a second limitation which is that all the downstream tasks used here were classification problems. Sequence learning problems (like for continuous speech transcription or for differentiating animal calls) or regression problems (like for medical diagnostics or emotion recognition) would be good targets for future work and might be better modelled by a different neural network architecture. Finally, future work could expand on these initial experiments by varying model size alongside training data size to quantify how pre-training data and model capacity interact. This study involved a large amount of pre-training data and was able to match pre-training dataset volumes across speech and non-speech domains. However, recent efforts have scaled dataset sizes beyond the volumes studied here (94k hours, [10] and beyond [46]), which could impact these results.

Training audio models without labels unlocks the potential of vast datasets that would otherwise be prohibitively expensive to annotate. SSL could drive data-discovery efforts (such as in urban planning research or medical applications), outlier detection (such as in monitoring for invasive species in bio-acoustics), or creative applications (such as using these representations to steer the synthesis of new sounds).

ACKNOWLEDGMENT

The author acknowledges the support from the Independent Research and Development (IRAD) Fund from the Johns Hopkins University Applied Physics Laboratory. Thanks to Han Yi, Will Coon and Lindsey Kitchell for helpful comments and discussion while drafting this manuscript. Feedback from ChatGPT (GPT-4o, OpenAI) was used to update some of the verbiage and language in this report.